\newcommand{\VINCI}{{\tt VINCI~}}
\newcommand{\POLYVEST}{{\tt PolyVest~}}

\documentclass[11pt]{article}

\usepackage[hmarginratio=1:1,top=32mm,columnsep=20pt]{geometry} 
\usepackage{hyperref} 
\usepackage{listings}
\usepackage{xcolor}

\usepackage{graphicx}

\usepackage{subfigure}
\usepackage{amsmath,amsthm,amssymb,amsfonts}
\usepackage{bm} 
\newtheorem{rem}{Theorem}

\usepackage{algorithm}
\usepackage{algpseudocode}

\usepackage{abstract} 

\title{\textbf{A Fast and Practical Method to Estimate Volumes of Convex Polytopes}}
\author{CunJing Ge, \ Feifei Ma, \ Jian Zhang \\ \\ \{gecj,maff,zj\}@ios.ac.cn \\Institute of Software, Chinese Academy of Sciences}

\date{}
\begin{document}
\maketitle
\thispagestyle{empty}

\begin{abstract}
The volume is an important attribute of a convex body.
In general, it is quite difficult to calculate the exact volume.
But in many cases, it suffices to have an approximate value.
Volume estimation methods for convex bodies have been extensively studied in theory,
however, there is still a lack of practical implementations of such methods.
In this paper, we present an efficient method which is based on the Multiphase Monte-Carlo algorithm to estimate volumes of convex polytopes.
It uses the coordinate directions hit-and-run method, and employs a technique of reutilizing sample points.
The experiments show that our method can efficiently handle instances with dozens of dimensions with high accuracy.\\[5pt]
\end{abstract}

\section{Introduction}
Volume computation is a classical problem in mathematics, arising in many appications such as economics, computational complexity analysis, linear systems modeling, and statistics. It is also extremely difficult to solve. Dyer et.al.~\cite{Dyer:1988} and Khachiyan~\cite{Khachiyan:1988, Khachiyan:1989}
proved respectively that exact volume computation is \#P-hard, even for explicitly described polytopes.
B{\"u}eler et.al.~\cite{Fukuda:2000} listed five volume computation algorithms for convex polytopes. However, only the instances around 10 dimensions can be solved in reasonable time with existing volume computation algorithms, which is quite insufficient in many circumstances. Therefore we turn attention to volume estimation methods.

There are many results about volume estimation algorithms of convex bodies since the end of 1980s.
A breakthrough was made by Dyer, Frieze and Kannan~\cite{Kannan:1989}.
They designed a polynomial time randomized approximation algorithm~(Multiphase Monte-Carlo Algorithm),
which was then adopted as the framework of volume estimation algorithms by successive works.
At first, the theoretical complexity of this algorithm is $O^*(n^{23})$
\footnote{``soft-O'' notation $O^*$ indicates that we suppress factors of $\log n$ as well as factors depending on other parameters like the error bound},
but it was soon reduced to $O^*(n^4)$ by Lov{\'a}sz, Simonovits et. al.~\cite{Lovasz:1990}\cite{Kannan:1996}\cite{Lovasz:1999}\cite{Lovasz:2006}. Despite the polynomial time results and reduced complexity, there is still a lack of practical implementation.
In fact, there are some difficulties in applying the above volume estimation algorithms.
First, in theoretical research of randomized volume algorithms, oracles are usually used to describe the convex bodies and the above time complexity results are measured in terms of oracle queries. 
However, oracles are too complex and oracle queries are time-consuming. Second, there exists a very large hidden constant coefficient in the theoretical complexity~\cite{Lovasz:1999}, which makes the algorithms almost infeasible even in low dimensions.
The reason leading to this problem is that the above research works mostly focus on arbitrary dimension and theoretical complexity.
To guarantee that Markov Chains mix in high-dimensional circumstance,
it is necessary to walk a large constant number of steps before determining the next point.

In this paper, we focus on practical and applicable method. We only consider specific and simple objects, i.e., convex polytopes. 
On the other hand, the size of problem instances is usually limited in practical circumstances.
With such limited scale, we find that it is unnecessary to sample as many points as the algorithm in~\cite{Lovasz:1999} indicates.
We implement a volume estimation algorithm which is based on the Multiphase Monte-Carlo method.
The algorithm is augmented with a new technique to reutilize sample points, so that the number of sample points can be significantly reduced.
We compare two hit-and-run methods: the hypersphere directions method and the coordinate directions method,
and find that the latter method which is employed in our approximation algorithm not only runs faster, but is also more accurate.
Besides, in order to better evaluate the performance of our tool, we also introduce a new result checking method.
Experiments show that our tool can efficiently handle instances with dozens of dimensions.
To the best of our knowledge, it is the first practical volume estimation tool for convex polytopes.

We now outline the remainder of the paper:
In section 2, we propose our method in detail.
In section 3, we show experimental results and compare our method with the exact volume computation tool {\tt VINCI}\cite{Vinciwebsite}.
Finally we conclude this paper in Section 4.

\section{The Volume Estimation Algorithm}
A convex polytope may be defined as the intersection of a finite number of half-spaces,
or as the convex hull of a finite set of points.
Accordingly there are two descriptions for a convex polytope:
half-space representation (H-representation) and vertex representation (V-representation).
In this paper, we adopt the H-representation.
An $n$-dimensional convex polytope $P$ is represented as $P = \{ Ax \le b \}$, where $A$ is an $(m \times n)$ matrix.
$a_{ij}$ represents the element at the $i$-th row and the $j$-th column of $A$,
and $a_i$ represents the $i$-th column vector of $A$.
For simplicity, we also assume that $P$ is full-dimensional and not empty.
We use $vol(K)$ to represent the volume of a convex body $K$, and $B(x, R)$ to represent the ball with radius $R$ and center $x$.

Like the original multiphase Monte-Carlo algorithm, our algorithm consists of three parts: rounding, subdivision and sampling.

\subsection{Rounding}
The rounding procedure is to find an affine transformation $T$ on polytope $Q$ such that $B(0, 1) \subseteq T(Q) \subseteq B(0, r)$
and a constant $\gamma=\frac{vol(Q)}{vol(T(Q))}$.
If $r > n$, $T$ can be found by the Shallow-$\beta$-Cut Ellipsoid Method~\cite{Book:1993}.
The Ellipsoid Method could take much time when $r$ is close to $n$, e.g. $r=n+1$.
There is a tradeoff between rounding and sampling,
since the smaller $r$ is, the more iterations during rounding and the fewer points have to be generated during sampling.
Rounding can handle very ``thin'' polytopes which cannot be subdivided or sampled directly.
We use $P$ to represent the new polytope $T(Q)$ in the sequel.
For more details about the rounding procedure, one can refer to Appendix~\ref{app:rounding}.

\subsection{Subdivision}
To avoid curse of dimensionality(the possibility of sampling inside a certain space in target object decreases very fast while dimension increases),
we subdivide $P$ into a sequence of bodies
so that the ratio of consecutive bodies is at most a constant, e.g. 2.
Place $l = \lceil n\log_2r \rceil$ concentric balls $\{B_i\}$ between $B(0, 1)$ and $B(0, r)$, where
\begin{displaymath}
B_i = B(0, r_i) = B(0, 2^{i/n}),\ i=0, \dots ,l.
\end{displaymath}
Set $K_i = B_i \cap P$, then $K_0 = B(0, 1)$, $K_l = P$ and
\begin{equation}\label{eqs1}
vol(P) = vol(B(0,1))\prod_{i=0}^{l-1} \frac{vol(K_{i+1})}{vol(K_i)} = vol(B(0, 1))\prod_{i=0}^{l-1} \alpha_i.
\end{equation}
So we only have to estimate the ratio $\alpha_i = vol(K_{i+1})/vol(K_i)$, $i=0, \dots ,l-1$.
Since $K_i = B_i \cap P \subseteq B_{i+1} \cap P = K_{i+1}$, we get $\alpha_i \ge 1$.
On the other hand, $\{K_i\}$ are convex bodies, then
\begin{displaymath}
K_{i+1} \subseteq \frac{r_{i+1}}{r_i}K_i = 2^{1/n}K_i,
\end{displaymath}
we have
\begin{displaymath}
\alpha_i = \frac{vol(K_{i+1})}{vol(K_i)} \le 2.
\end{displaymath}
Specially, $K_{i+1} = 2^{1/n}K_i$ if and only if $K_{i+1} = B_{i+1}$ i.e. $B_{i+1} \subseteq P$.
That is, $1\le\alpha_i\le 2$ and $\alpha_i = 2 \Leftrightarrow B_{i+1} \subseteq P$.

\subsection{Hit-and-run}
To approximate $\alpha_i$, we generate $step\_size$ random points in $K_{i+1}$ and count the number of points $c_i$ in $K_i$.
Then $\alpha_i \approx step\_size / c_i$.
It is easy to generate uniform distributions on cubes or ellipsoids but not on $\{K_i\}$.
So we use a random walk method for sampling.
Hit-and-run method is a random walk which has been proposed and studied for a long time~\cite{Smith:1984}\cite{Smith:1987}\cite{Belisle:1993}.
The hypersphere directions method and the coordinate directions method are two hit-and-run methods.
In the hypersphere directions method, the random direction is generated from a uniform distribution on a hypersphere;
in the coordinate directions method, it is chosen with equal probability from the coordinate direction vectors and their negations.
Berbee et al.~\cite{Smith:1987} proved the following theorems.
\begin{rem}\label{rem:hypersphere}
The hypersphere directions algorithm generates a sequence of interior points whose limiting distribution is uniform.
\end{rem}
\begin{rem}\label{rem:coordinate}
The coordinate directions algorithm generates a sequence of interior points whose limiting distribution is uniform.
\end{rem}
Coordinate directions and their negations are special cases of directions generated on a hypersphere,
hence the former theoretical research about volume approximation algorithm with hit-and-run methods mainly focus on the hypersphere directions method~\cite{Lovasz:1999}.
In this paper, we apply the coordinate directions method to our volume approximation algorithm.
We will compare practical performances of two methods in Section~\ref{section:comp}.

\subsection{Reutilization of Sample Points}
In the original description of the Multiphase Monte Carlo method,
it is indicated that the ratios $\alpha_i$ are estimated in natural order,
from the first ratio $\alpha_0$ to the last one $\alpha_{l-1}$. 
The method starts sampling from the origin.
At the $k$th phase, it generates a certain number of random independent points in $K_{k+1}$ and counts the number of points $c_k$ in $K_k$ to estimate $\alpha_k$.
However, our algorithm performs in the opposite way:
Sample points are generated from the outermost convex body $K_l$ to the innermost convex body $K_0$,
and ratios are estimated accordingly in reverse order.

The advantage of approximation in reverse order is that it is possible to fully exploit the sample points generated in previous phases.
Suppose we have already generated a set of points $\mathcal{S}$ by random walk with almost uniform distribution in $K_{k+1}$,
and some of them also hit the convex body $K_k$, denoted by $\mathcal{S}'$.
The ratio $\alpha_k$ is thus estimated with $\frac{|\mathcal{S}'|}{|\mathcal{S}|}$.
But these sample points can reveal more information than just the ratio $\alpha_k$.
Since $K_k$ is a sub-region of $K_{k+1}$, the points in $\mathcal{S}'$ are also almost uniformly distributed in $K_k$.
Therefore, $\mathcal{S}'$ can serve as part of the sample points in $K_k$.
Furthermore, for any $K_i$ ($0 \le i \le k$) inside $K_{k+1}$,
the points in $K_{k+1}$ that hit $K_i$ can serve as sample points to approximate $\alpha_i$ as well. 

Based on this insight, our algorithm samples from outside to inside.
Suppose to estimate each ratio within a given relative error, we need as many as $step\_size$ points.
At the $k$th phase which approximates ratio $\alpha_{l-k}$,
the algorithm first calculates the number $count$ of the former points that are also in $\alpha_{l-k+1}$,
then generates the rest $(step\_size-count)$ points by random walk.  

Unlike sampling in natural order, choosing the starter for each phase in reverse sampling is a bit complex. The whole sampling process in reverse order also starts from the origin point.
At each end of the $k$-th phase, we select a point $x$ in $K_{k+1}$
and employ $x' = 2^{-\frac{1}{n}}x$ as the starting point of the next phase since $2^{-\frac{1}{n}}x \in K_k$.

It's easy to find out that the expected number of reduced sample points with our algorithm is  
\begin{equation}\label{eqs4}
\sum_{i=1}^{l-1} (step\_size \times \frac{1}{\alpha_i}).
\end{equation}

Since $\alpha_i \le 2$, we only have to generate less than half sample points with this technique.
Actually, results of expriments show that we can save over 70\% time consumption on many polytopes.

\subsection{Framework of the Algorithm}
Now we present the framework of our volume estimation method. Algorithm~\ref{alg:main} is the Multiphase Monte-Carlo algorithm with the technique of reutilizing sample points.

\begin{algorithm}[!htbp]
\caption{The Framework of Volume Estimation Algorithm}\label{alg:main}
\begin{algorithmic}[1]
\Function{EstimateVol}{}
	\State $\gamma \leftarrow Preprocess(\,)$
	\State $x \leftarrow O$
	\State $l \leftarrow \lceil n\log_2r \rceil$
	\For{$k \leftarrow l-1, \ 0$}
		\For{$i \leftarrow count,\ step\_size$}
			\State $x \leftarrow Walk(x, k)$
			\If{$x \in B_0$}
				\State $t_0 \leftarrow t_0 + 1$
			\ElsIf {$x \in B_k$}
				\State $m \leftarrow \lceil \frac{n}{2}\log_2|x| \rceil$
				\State $t_{m} \leftarrow t_{m} + 1$
			\EndIf
		\EndFor
		\State $count \leftarrow \sum_{i=0}^{k} t_i$
		\State $\alpha_k \leftarrow step\_size / count$
		\State $x \leftarrow 2^{-\frac{1}{n}}x$
	\EndFor
	\State \textbf{return} $\gamma \cdot unit\_ball(n) \cdot \prod_{i=0}^{l-1} \alpha_i$
\EndFunction
\end{algorithmic}
\end{algorithm}

In Algorithm~\ref{alg:main}, the formula $\lceil \frac{n}{2}\log_2|x| \rceil$ returns index $i$ that $x \in K_i \setminus K_{i-1}$.
We use $t_i$ to record the number of sample points that hit $K_i \setminus K_{i-1}$.
Furthermore, the sum $count$ of $t_0, \dots, t_{k+1}$ is the number of reusable sample points that are generated inside $K_{k+1}$.
Then we only have to generate the rest $(step\_size-count)$ points inside $K_{k+1}$ in the $k$-th phase.
Then we use $2^{-\frac{1}{n}}x$ as the starting point of the next phase.
Finally, according to equation (\ref{eqs1}) and $\gamma=\frac{vol(Q)}{vol(P)}$, we achieve the estimation of $vol(Q)$ .

\section{Experimental Results}
We implement the algorithm in C++ and the tool is named \POLYVEST (Polytope Volume Estimation). 
In all experiments, $step\_size$ is set to $1600l$ for the reason discussed in Appendix~\ref{app:err}
and parameter $r$ is set to $2n$.
The experiments are performed on a workstation with 3.40GHz Intel® Core™ i7-2600 CPU and 8GB memory. Both \POLYVEST and \VINCI use a single core.

\subsection{The Performance of \POLYVEST}
Table~\ref{table:comp} shows the results of comparison between \POLYVEST and {\tt VINCI}.
\VINCI is a well-known package which implements the state of the art algorithms for exact volume computation of convex polytopes.
It can accept either H-representation or V-representation as input. 
The test cases include:
(1) ``cube\_n'': Hypercubes with side length $2$, i.e. the volume of ``cube\_n'' is $2^n$.
(2) ``cube\_n(S)'': Apply 10 times random shear mappings on ``cube\_n''.
The random shear mapping can be represented as  
$PQP$, with $Q=\left(\begin{array}{cc}I & M \\0 & I\end{array}\right)$,
where the elemets of matrix $M$ are randomly chosen
and $P$ is the products of permutation matrices $\{P_i\}$ that put rows and columns of $Q$ in random orders.
This mapping preserves the volume.
(3) ``rh\_n\_m'': An $n$-dimentional polytope constructed by randomly choosing $m$ hyperplanes tangent to sphere.
(4) ``rh\_n\_m(S)'': Apply 10 times random shear mappings on ``rh\_n\_m''.
(5) ``cuboid\_n(S)'': Scaling ``cube\_n'' by 100 in one direction, and then apply random shear mapping on it once.
We use this instance to approximate a ``thin stick'' which not parallel to any axis.
(6) ``ran\_n\_m'': An $n$-dimentional polytope constructed by randomly choosing integer coefficient from -1000 to 1000 of matrix $A$.

\begin{table}[!htbp]
\footnotesize \centering
\setlength{\abovecaptionskip}{0pt}
\setlength{\belowcaptionskip}{5pt}
\caption{Comparison between \POLYVEST and \VINCI}\label{table:comp}
\begin{tabular}{|c|c|c|c|c|c|c|c|c|}
\hline
\multicolumn{3}{|c|}{} & 	\multicolumn{2}{|c|}{\POLYVEST} & \multicolumn{4}{|c|}{\VINCI} \\
\hline
Instance	& $n$	& $m$	& Result		& Time(s)	& Result	& $T_{rlass}$(s)	& $T_{hot}$(s)	& $T_{lawnd}$(s)	\\
\hline
cube\_10	& 10	& 20	& 1015.33		& 0.380		& 1024		& 0.004		& 0.044		& 0.008	\\
cube\_15	& 15	& 30	& 33560.1		& 1.752		& 32768		& 0.300 	& 212.8		& 0.156 \\
cube\_20	& 20	& 40	& 1.08805e+6	& 4.484		& 1.04858e+6	& ---	& ---		& 8.085	\\
cube\_30	& 30	& 60	& 1.0902e+9		& 23.197	& ---		& ---		& ---		& ---	\\
cube\_40	& 40	& 80	& 1.02491e+12	& 72.933	& ---		& ---		& ---		& ---	\\
cube\_10(S)	& 10	& 20	& 1027.1		& 0.184		& 1023.86	& 0.008		& 0.124		& 0.024 \\
cube\_15(S)	& 14	& 28	& 30898.2		& 0.784		& 32766.4	& 0.428		& 369.6		& 0.884 \\
rh\_8\_25	& 8		& 25	& 793.26		& 0.132		& 785.989	& 0.864		& 0.160		& 0.016 \\
rh\_10\_20	& 10	& 20	& 13710.0		& 0.240		& 13882.7	& 0.284		& 0.340		& 0.012 \\
rh\_10\_25	& 10	& 25	& 5934.99		& 0.260		& 5729.52	& 5.100		& 1.932		& 0.072 \\
rh\_10\_30	& 10	& 30	& 2063.55		& 0.280		& 2015.58	& 660.4*	& 5.772		& 0.144 \\
rh\_8\_25(S)	& 8		& 25	& 782.58	& 0.136		& 785.984	& 1.268		& 0.156		& 0.032 \\
rh\_10\_20(S)	& 10	& 20	& 13773.2	& 0.232		& 13883.8	& 0.832		& 0.284		& 0.032 \\
rh\_10\_25(S)	& 10	& 25	& 5667.49	& 0.252		& 5729.18	& 11.949	& 1.960		& 0.104 \\
rh\_10\_30(S)	& 10	& 30	& 2098.89	& 0.276		& 2015.87	& 1251.1*	& 6.356		& 0.248 \\
\hline
\end{tabular}\\
*: Enable the \VINCI option to restrict memory storage, so as to avoid running out of memory.
\end{table}

In Table~\ref{table:comp}, $T_{rlass}$, $T_{hot}$ and $T_{lawnd}$ represent the time consumption of three parameters of methods in \VINCI respectively.
The ``rlass'' uses Lasserre's method, it needs input of H-representation.
The ``hot'' uses a Cohen\&Hikey-like face enumeration scheme, it needs input of V-representation.
The ``lawnd'' uses Lawrence's formula, it is the fatest method in \VINCI and both descriptions are needed.
From ``cube\_20'' to ``cube\_40'', ``rlass'' and ``hot'' cannot handle these instances in reasonable time.
We did not test instances ``cube\_30'' and ``cube\_40'' by ``lawnd'', because there are too many vertices in these polytopes.

Observe that the ``rlass'' and ``hot'' methods of \VINCI usually take much more time and space as the scale of the problem grows a bit,
e.g. ``cube\_n($n\ge 15$)'' and ``rh\_10\_30''.
With H- and V- representations, the ``lawnd'' method is very fast for instances smaller than 20 dimensions.
However, enumerating all vertices of polytopes is non-trivial, as is the dual problem of constructing the convex hull by the vertices.
Such process is either time-consuming and space-consuming
that makes ``lawnd'' method slower than \POLYVEST for random polytopes around 15 dimensions which only given by hyperplanes.
The running times of \POLYVEST appear to be more `stable'.
In addition, \POLYVEST only has to store some constant matrices and variable vectors for sampling.

\begin{table}[!htbp]
\footnotesize \centering
\setlength{\abovecaptionskip}{0pt}
\setlength{\belowcaptionskip}{5pt}
\caption{Statistical Results of \POLYVEST}\label{table:results1}
\begin{tabular}{|c|c|c|c|c|c|}
\hline
Instance 		& Average	& Std Dev  		& $95\%$ Confidence Interval & Freq	& Error	\\
		 		& Volume $\overline{v}$ 	& $\sigma$ &
$\mathcal{I}=[p, q]$ & on $\mathcal{I}$ 	& $\epsilon = \frac{q - p}{\overline{v}}$	\\
\hline
cube\_10*		& 1024.91		& 41.7534		& [943.077, 1106.75]		& 947	& 15.9695\%	\\
cube\_20*		& 1.04551e+6	& 49092.6		& [9.49284e+5, 1.14173e+6] 	& 942	& 18.4067\%	\\
cube\_30		& 1.06671e+9	& 5.95310e+7 	& [9.50024e+8, 1.18339e+9]	& 96	& 21.8769\%	\\
cube\_40		& 1.09328e+12	& 4.85772e+10 	& [9.98073e+11, 1.18850e+12] & 95	& 17.4175\%	\\
cuboid\_10(S)* 	& 102258		& 3162.13 		& [96060.1, 108456]			& 953	& 12.1219\%	\\
cuboid\_20(S)* 	& 1.04892e+8	& 388574e+6		& [9.72760e+7, 1.12508e+8]	& 953	& 14.5217\%	\\
cuboid\_30(S) 	& 1.07472e+11	& 4.42609e+9 	& [9.87968e+10, 1.16147e+11] & 93	& 16.1440\%	\\
ran\_10\_30* 	& 11.0079		& 0.413874		& [10.1967, 11.8191]		& 946	& 14.7383\%	\\
ran\_10\_50* 	& 1.48473		& 4.81726e-2	& [1.39031, 1.57915]		& 952	& 12.7186\%	\\
ran\_15\_30 	& 290.575		& 12.8392		& [265.410, 315.740]		& 92	& 17.3208\% \\
ran\_15\_50 	& 3.30084		& 0.145495		& [3.01567, 3.58601]		& 96	& 17.2787\% \\
ran\_20\_50 	& 1.25062		& 6.60574e-2	& [1.12115, 1.38010]		& 94	& 20.7053\% \\
ran\_20\_100 	& 8.79715e-3	& 3.144633e-4	& [8.18080e-3, 9.41350e-3]	& 96	& 14.0125\% \\
ran\_30\_60 	& 195.295		& 10.37041		& [174.969, 215.621]		& 97	& 20.8157\%	\\
ran\_30\_100 	& 2.21532e-5	& 1.13182e-6	& [1.99348e-5, 2.43715e-5]	& 98	& 20.0276\% \\
ran\_40\_100 	& 3.02636e-5	& 1.76093e-6	& [2.68121e-5, 3.3715e-5]	& 96	& 22.8091\%	\\
\hline
\end{tabular}\\
*: Estimated 1000 times with {\tt POLYVEST}.
\end{table}

Since \POLYVEST is a volume estimation method instead of an exact volume computation one like {\tt VINCI},
we did more tests on \POLYVEST to see how accurate it is.
We estimated 100 times with \POLYVEST for each instance in Table~\ref{table:results1} and listed the statistical results.
From Table~\ref{table:results1}, we observe that the frequency on $\mathcal{I}$ is approximately 950
which means $Pr(p \le \overline{vol(P)} \le q) \approx 0.95$.
Additionally, values of $\epsilon$~(ratio of confidence interval's range to average volume $\overline{v}$)
are smaller than or around 20\%.

\subsection{Result Checking}
For arbitrary convex polytopes with more than 10 dimensions,
there is no easy way to evaluate the accuracy of \POLYVEST since the exact volumes cannot be computed with tools like {\tt VINCI}.
However, we find that a simple property of geometric body is very helpful for verifying the results.   

Given an arbitrary geometric body $P$, an obvious relation is that if $P$ is divided into two parts $P_1$ and $P_2$,
then we have $vol(P)= vol(P_1)+vol(P_2)$.
For a random convex polytope, we randomly generate a hyperplane to cut the polytope,
and test if the results of \POLYVEST satisfy this relation.

Table~\ref{table:result2} shows the results of such tests on random polytopes in different dimensions.
Each polytope is tested 100 times.
Values in column ``Freq.'' are the times that $(vol(P_1)+vol(P_2))$ falls in 95\% confidence interval of $vol(P)$,
and these values are all greater than 95. The error $\frac{\mid Sum-\overline{vol(P)}\mid}{\overline{vol(P)}}$ is quite small.
Therefore, the outputs of \POLYVEST satisfy the relation $vol(P)= vol(P_1)+vol(P_2)$. The test results further confirm the reliability of {\tt PolyVest}. 

\begin{table}[!htbp]
\footnotesize \centering
\setlength{\abovecaptionskip}{0pt}
\setlength{\belowcaptionskip}{5pt}
\caption{Result Checking}\label{table:result2}
\begin{tabular}{|c|c|c|c|c|c|c|c|}
\hline
$n$	& $\overline{vol(P)}$	& 95\% Confidence Interval	& $\overline{vol(P_1)}$	& $\overline{vol(P_2)}$ & Sum & Error	& Freq. \\
\hline
10	& 916.257	& [847.229, 985.285]	& 498.394	& 414.676	& 913.069	& 0.348\%	& 98	\\
20	& 107.976	& [97.4049, 118.548]	& 50.4808	& 57.3418	& 107.823	& 0.142\%	& 99	\\
30	& 261424	& [228471, 294376]		& 40332.7	& 218637	& 258969	& 0.939\%	& 96	\\
40	& 5.07809e+11 & [4.58326e+11, 5.57292e+11] & 9.43749e+10 & 4.14623e+11 & 5.08997e+11 & 0.234\% & 98	\\
\hline
\end{tabular}
\end{table}

\subsection{The Performance of two Hit-and-run Method}\label{section:comp}
In Table~\ref{table:hr1}, $t_1$ and $t_2$ represent the time consumption of the coordinate directions and the hypersphere directions method
when each method is executed 10 million times.
Observe that the coordinate directions method is faster than the other one.
The reason is that the hypersphere directions method has to do more vector multiplications to find intercestion points
and $m\times n$ more divisions during each walk step.

\begin{table}[!htbp]
\footnotesize \centering
\setlength{\abovecaptionskip}{0pt}
\setlength{\belowcaptionskip}{5pt}
\caption{Random walk by 10 million steps}\label{table:hr1}
\begin{tabular}{|c|c|c|c|}
\hline
$n$	& $m$	& time $t_1$(s)	& time $t_2$(s) \\
\hline
10	& 20	& 6.104		& 13.761 \\
20	& 40	& 10.701	& 24.502 \\
30	& 60	& 17.541	& 40.455 \\
40	& 80	& 27.494	& 61.484 \\
\hline
\end{tabular}
\end{table}

In addition, we also compare the two hit-and-run methods on accuracy.
The results in Table~\ref{table:hr2} show that the relative errors and standard deviations of the coordinate directions method are smaller.

\begin{table}[!htbp]
\footnotesize \centering
\setlength{\abovecaptionskip}{0pt}
\setlength{\belowcaptionskip}{5pt}
\caption{Comparison about accuracy between two methods}\label{table:hr2}
\begin{tabular}{|c|c|c|c|c|c|c|c|c|}
\hline
\multicolumn{2}{|c|}{} & 	\multicolumn{3}{|c|}{Simplified} & \multicolumn{3}{|c|}{Original} \\
\hline
Instance	& Exact Vol $v$	& Volume $\overline{v}$ & Err $\frac{\mid\overline{v}-v\mid}{v}$	& Std Dev $\sigma$
	& Volume $\overline{v}'$	& Err $\frac{\mid\overline{v}-v\mid}{v}$	& Std Dev $\sigma'$	\\
\hline
cube\_10	& 1024			& 1024.91		& 0.089\%	& 41.7534	& 1028.31		& 0.421\%	& 62.6198	\\
cube\_14	& 16384			& 16382.3		& 0.010\%	& 3.020		& 16324.6		& 0.363\%	& 1145.76	\\
cube\_20	& 1.04858e+6	& 1.04551e+6	& 0.293\%	& 49092.6	& 1.04426e+6	& 0.412\%	& 81699.9	\\
rh\_8\_25	& 785.989		& 786.240		& 0.032\%	& 23.5826	& 791.594		& 0.713\%	& 50.5415	\\
rh\_10\_20	& 13882.7		& 13876.3		& 0.046\%	& 473.224	& 13994.4		& 0.805\%	& 963.197	\\
rh\_10\_25	& 5729.52		& 5736.83		& 0.128\%	& 193.715	& 5765.18		& 0.622\%	& 368.887	\\
rh\_10\_30	& 2015.58		& 2013.08		& 0.124\%	& 62.1032	& 2041.60		& 1.291\%	& 124.204	\\
\hline
\end{tabular}
\end{table}

\subsection{The Advantage of Reutilization of Sample Points}
In Table~\ref{table:result_re}, we demonstrate the effectiveness of reutilization technique.
Values of $n_1$ are the number of sample points without this technique.
Since our method is a randomized algorithm, the number of sample points with this technique is not a constant.
So we list average values in column $n_2$.
With this technique, the requirement of sample points is significantly reduced.

\begin{table}[!htbp]
\footnotesize \centering
\setlength{\abovecaptionskip}{0pt}
\setlength{\belowcaptionskip}{5pt}
\caption{Reutilize Sample Points}\label{table:result_re}
\begin{tabular}{|c|c|c|c|}
\hline
Instance	& $n_1$ 	& $n_2$		& $n_2/n_1$ \\
\hline
cube\_10	& 2016000	& 535105.41	& 26.5\%	\\
cube\_15	& 5856000	& 1721280.3	& 29.4\%	\\
cube\_20	& 12249600	& 3789370.7	& 30.9\%	\\
rh\_8\_25	& 1040000	& 181091.13	& 17.4\%	\\
rh\_10\_30	& 2016000	& 304211.03	& 15.1\%	\\
cross\_7	& 809600	& 78428.755	& 9.69\%	\\
fm\_6		& 5856000	& 955656.79	& 16.3\%	\\
\hline
\end{tabular}
\end{table}

\section{Related Works}
To our knowledge, there are only two implementations of volume estimation methods in literature.
Liu et al.~\cite{Liu:2007} developed a tool to estimate volume of convex body with a direct Monte-Carlo method.
Suffered from the curse of dimensionality,
it can hardly solve problems as the dimension reaches 5.
The recent work~\cite{Lovasz:2012} is an implementation of the $O^*(n^4)$ volume algorithm in~\cite{Lovasz:2006}.
Some interesting techniques are also discussed in the paper.
However, the algorithm is targeted for convex bodies,
and only the computational results for instances within 10 dimensions are reported.
The authors also report that they could not experiment with other convex bodies than cubes,
since the oracle describing the convex bodies took too long to run.

\section{Conclusion}
In this paper, we propose an efficient volume estimation algorithm for convex polytopes
which is based on Multiphase Monte Carlo algorithm.
With simplified hit-and-run method and the technique of reutilizing sample points,
we considerably improve the existing algorithm for volume estimation and implement a practical tool.
Our tool, {\tt PolyVest}, can efficiently handle instances with dozens of dimensions with high accuracy,
while the exact volume computation algorithms often fail on instances with over 10 dimensions.
In fact, the complexity of our method~(excluding rounding procedure) is $O^*(mn^3)$
and it is measured in terms of basic operations instead of oracle queries.
Therefore, our method requires much less computational overhead than the theoretical algorithms.

\section{Acknowledgement}
The authors would like to thank Peng Zhang for his comments and suggestions, and Xingming Wu for being helpful in evaluating the tool.

\begin{appendix}

\section{Rounding}\label{app:rounding}
The pseudocode of rounding procedure and other preprocessings is presented in Algorithm~\ref{alg:prep}.
We define ellipsoid $E = \{x \in \mathbb{R}^n | (x-a)^TA^{-1}(x-a) \le 1\}$,
where $A$ is a symmetric positive definite matrix.
In function $InitEllipsoid$, we maximize each of the $2n$ linear functions $x_1, -x_1,  \dots , x_n, -x_n$ subject to $Ax \le b$.
So we get bounds ${UB_1, LB_1,  \dots , UB_n, LB_n}$ of each dimension of $P$ and $2n$ vertices $v_1, \dots ,v_{2n}$
(possible that $v_i = v_j, i \not= j$).
Let $o_0 = \frac{1}{2n} \sum_{i=1}^{2n}v_i$ and $r_0 = \sqrt{\sum_{i=1}^{n}(UB_i-LB_i)^2}$.
Then we obtain the initial ellipsoid $E_0(r_0^2I, o_0) = B(o_0, r_0)$
where $o_0 \in P$(notice that $P$ is a convex body) and $P \subseteq E_0$.

Line 3--20 of Algorithm~\ref{alg:prep} is the implementation of Shallow-$\beta$-Cut Ellipsoid Method~\cite{Book:1993}.
It is an iterative method that generates a series of ellipsoids $\{E_i(T_i, o_i)\}$ s.t. $P \subseteq E_i$,
until we find an $E_k$ such that $E_k(\beta^2 T_k, o_k) \subseteq P$, where $\beta = \frac{1}{r}$ and $0<\beta<1/n$.

\begin{algorithm}[!htbp]
\caption{The Ellipsoid Method and the affine transformation}\label{alg:prep}
\begin{algorithmic}[1]
\Function{Preprocess}{}
	\State $InitEllipsoid(r_0, o_0)$
	\State $T_0 \leftarrow r_0^2 \cdot I$
	\State $k \leftarrow 0$
	\Loop
		\State $i \leftarrow -1$
		\If{$o_k \notin P$}
			\State $choose\ i\ that\ a_ix \le b_i\ does\ not\ hold$
		\ElsIf{$E(\beta^2 T_k, o_k) \nsubseteq P$}
			\State $choose\ i\ such \ that\ \beta^2 a_iT_ka_i^T \le (b_i-a_io_k)\ does\ not\ hold$
		\EndIf
		\If{$i \ge 0$}
			\State $c \leftarrow \frac{T_ka_i^T}{\sqrt{a_iT_ka_i^T}}$
			\State $o_{k+1} \leftarrow o_k - \frac{1-n\beta}{n+1}c^T$
			\State $T_{k+1} \leftarrow (1+\frac{(1-n\beta)^2}{2n^2})\frac{n^2(1-\beta^2)}{n^2-1}(T_k-\frac{2(1-n\beta)cc^T}{(n+1)(1-\beta)})$
		\Else
		\State $break\ loop$
		\EndIf
		\State $k \leftarrow k+1$
	\EndLoop
	\State $L \leftarrow Cholesky(T_k)$
	\State $b \leftarrow (b - Ao_k) / \beta$
	\State $A \leftarrow AL^T$
	\State \textbf{return} $det(L)\beta^n$
\EndFunction
\end{algorithmic}
\end{algorithm}

The affine transformation is described through Line 21-24.
Function $Cholesky(T_k)$ returns the Cholesky factorization $L$ of $T_k$ (that is, $T_k = L^TL$ and $L$ is an upper triangular matrix),
since $T_k$ is a symmetric positive definite matrix.
Notice
\begin{displaymath}
E_k(T_k,o_k) = E_k(L^TL, o_k) = \{x \in \mathbb{R}^n | ((L^T)^{-1}(x-o_k))^T(L^T)^{-1}(x-o_k) \le 1\}.
\end{displaymath}
Let $y = (L^T)^{-1}(x-o_k)$,
then $\{y \in \mathbb{R}^n | y^Ty \le 1\} = B(0, 1)$. Thus
\begin{displaymath}
E_k(T_k, o_k) = L^TB(0, 1) + o_k.
\end{displaymath}
Substitute $x$ in $P = \{Ax \le b\}$ by $x = L^Ty+o_k$, we get
\begin{equation}\label{eqs5}
P' = \{A(L^Ty+o_k) \le b\} = \{A'y \le b'\},\ B(0,\beta) \subseteq P' \subseteq B(0, 1),
\end{equation}
where $A' = AL^T$, $b' = b - Ao_k$.\\
Resize $P'$ by ratio $\frac{1}{\beta}$, $B(0, 1) \subseteq P'' = \frac{1}{\beta}P' \subseteq B(0, \frac{1}{\beta})$
\begin{equation}\label{eqs6}
where\ P'' = \{A''x \le b''\},\ A'' = AL^T,\ b'' = \frac{b - Ao_k}{\beta}.
\end{equation}

The formulas in (\ref{eqs6}) are that of line 22, 23 in Algorithm~\ref{alg:prep}.
From (\ref{eqs5}) and (\ref{eqs6}),
\begin{equation}\label{eqs7}
\gamma = \frac{vol(P)}{vol(P'')} = det(L)\beta^n.
\end{equation}
So in Algorithm~\ref{alg:prep}, function $Preprocess$ returns the ratio of $\gamma$.

\section{About the Number of Sample Points}\label{app:err}
From Formula (\ref{eqs1}),
\begin{displaymath}
\frac{vol(P)}{vol(B(0,1))} = \prod_{i=0}^{l-1} \alpha_i =
\prod_{i=0}^{l-1} \frac{step\_size}{c_i} = \frac{step\_size^l}{\prod_{i=0}^{l-1} c_i},
\end{displaymath}
which shows that to obtain confidence interval of $vol(P)$, we only have to focus on $\prod_{i=0}^{l-1} c_i$.
For a fixed $P$, $\{\alpha_i\}$ are fixed numbers.
Let $c = \prod_{i=1}^{l}c_i$ and $\mathbb{D}(l, P)$ denote the distribution of $c$.
With statistical results of substantial expriments on concentric balls,
we observe that, when $step\_size$ is sufficiently large, the distribution of $c_i$ is unbiased
and its standard deviation is smaller than twice of the standard deviation of binomial distribution in dimensions below 80.
Though such observation sometimes not holds when we sample on convex bodies other than balls,
we still use this to approximate the distribution of $c_i$.
Consider random variables $X_i$ following binomial distribution $\mathbb{B}(step\_size, 1 / \alpha_i)$, we have
\begin{displaymath}
E(c) = E(c_1) \dots E(c_l) = E(X_1) \dots E(X_l) = step\_size^l \prod_{i=1}^{l}\frac{1}{\alpha_i},
\end{displaymath}
\begin{eqnarray*}
D(c) & = & E((c_1 \dots c_l)^2) - E(c)^2 = \prod_{i=1}^{l}(D(c_i)+E(c_i)^2) - E(c)^2\\
	 & = & \prod_{i=1}^{l}(4D(X_i)+E(X_i)^2) - E(c)^2\\
	 & = & \prod_{i=1}^{l}\frac{step\_size^2}{\alpha_i^2}(1+\frac{4\alpha_i}{step\_size}(1-\frac{1}{\alpha_i})) - E(c)^2\\
	 & = & E(c)^2(\beta - 1),
\end{eqnarray*}
where $\beta = \prod_{i=1}^{l}(1 + \frac{4\alpha_i}{step\_size} - \frac{4}{step\_size})$.

Suppose $\{\xi_1,  \dots , \xi_{t}\}$ is a sequence of i.i.d. random variables following $\mathbb{D}(l, P)$.
Notice $D(c)$, the variance of $\mathbb{D}(l, P)$,
is finite because $\beta - 1 \rightarrow 0$ as $t\rightarrow\infty$.
According to \textbf{central limit theorem}, we have
\begin{displaymath}
\frac{\sum_{i=1}^{t}\xi_i - tE(c)}{\sqrt{t} D(c)} \stackrel{d}{\rightarrow} N(0, 1).
\end{displaymath}
So we obtain the approximation of $95\%$ confidence interval of $c$,
$[E(c) - \sigma\sqrt{D(c)}, E(c) + \sigma\sqrt{D(c)}]$, where $\sigma = 1.96$.
And
\begin{displaymath}
Pr(\frac{vol(B(0, 1))step\_size^l}{E(c) + \sigma\sqrt{D(c)}} \le
\overline{vol(P)} \le \frac{vol(B(0, 1))step\_size^l}{E(c) - \sigma\sqrt{D(c)}}) \approx 0.95.
\end{displaymath}
Let $\epsilon \in [0, 1]$ denote the ratio of confidence interval's range to exact value of $vol(P)$, that is
\begin{eqnarray}
& & \frac{vol(B(0, 1))step\_size^l}{E(c) + \sigma\sqrt{D(c)}} -
\frac{vol(B(0, 1))step\_size^l}{E(c) - \sigma\sqrt{D(c)}} \le vol(P) \cdot \epsilon\\
& \Longleftrightarrow & \frac{1}{E(c) - \sigma\sqrt{D(c)}} -
\frac{1}{E(c) + \sigma\sqrt{D(c)}} \le \frac{\epsilon}{E(c)}\\
& \Longleftrightarrow & \frac{1}{1 - \sigma\sqrt{\beta - 1}} - \frac{1}{1 + \sigma\sqrt{\beta - 1}} \le \epsilon\\
& \Longleftrightarrow & 4\sigma^2(\beta - 1) \le \epsilon^2 (1 + \sigma^2 - \sigma^2\beta)^2\\
& \Longleftrightarrow & \epsilon^2\sigma^2\beta^2 - 2\epsilon^2(1+\sigma^2)\beta - 4\beta + (\frac{1}{\sigma}+\sigma)^2 + 4 \ge 0.\label{eqs12}
\end{eqnarray}
Solve inequality (\ref{eqs12}),
we get $\beta_1(\epsilon, \sigma)$, $\beta_2(\epsilon, \sigma)$ that $\beta \le \beta_1$ and $\beta \ge \beta_2$
(ignore $\beta \ge \beta_2$ because $1 - \sigma \sqrt{\beta_2-1} < 0$).
$\beta \le (1 + \frac{4}{step\_size})^l$, since $1 \le \alpha_i \le 2$.
\begin{equation}
(1 + \frac{4}{step\_size})^l \le \beta_1 \Longleftrightarrow step\_size \ge \frac{4}{\beta_1^{1 / l} - 1},\label{eqs13}
\end{equation}
(\ref{eqs13}) is a sufficient condition of $\beta \le \beta_1$.
Furthermore, $4 / (l\beta_1^{1 / l} - l)$ is nearly a constant as $\epsilon$ and $\sigma$ are fixed.
For example, $4 / (l\beta_1^{1 / l} - l) \approx 1569.2 \le 1600$ when $\epsilon = 0.2$, $\sigma = 1.96$.
So $step\_size = 1600l$ keeps the range of $95\%$ confidence interval of $vol(P)$ less than $20\%$ of the exact value of $vol(P)$.

\end{appendix}

\end{document}